\documentclass{article}
\usepackage{graphicx}
\begin{document}
\title{\bf{Summary of the theory of noncontractible physical space}}
\author{\bf{Davor Palle} \\
ul. Ljudevita Gaja 35, 10000 Zagreb, Croatia \\
email: davor.palle@gmail.com}
\maketitle
\begin{abstract}
{
There are a plenty of phenomenological, mathematical and philosophical
reasons to modify or generalize our currently accepted theories
in particle physics, cosmology and gravity.
I give a comprehensive list of anomalies, puzzles and problems
in the field of particle physics, cosmology and gravity that is
the essential motivation for the curiosity driven research.
The critiques of the Higgs mechanism, inflationary mechanism, instantons,
sphalerons, gravity waves and the Hawking radiation are given in details.
I introduce the minimal distance (UV cut-off) to resolve the
UV singularity problem in quantum field theory of particle physics,
minimal distance in quantum mechanics and in the nonsingular Einstein-Cartan
theory of gravity relevant in cosmology.
The universality of the minimal distance can be and should be checked
within particle physics, astrophysics and cosmology.
The dark matter problem, the dark energy problem, light neutrino masses and mixing,
cosmic rays muon puzzle, B meson anomalies, Tevatron t-quark charge asymmetry,
W boson mass anomaly, lepton and baryon number violations, chirality of the vorticity of the Universe,
kinematic dipole anomaly, etc., all have unique and clear solutions within my theory.
The dimensionality of spacetime and a perfect balance between the translational
and rotational degrees of freedom in Minkowski and Riemann-Cartan spacetimes are the crucial
ingredients in my reasoning.
}
\end{abstract}

\section{Anomalies, puzzles, tensions and problems as a motivation for research}
I divide the review of the main obstacles in physics into three domains: (A) particle physics,
(B) cosmology and (C) gravity.

{\bf (A) particle physics - Standard Model}

Our current understanding of the world of particle physics is completely described by the Standard Model (SM)
containig the QCD $SU(3)$ gauge symmetry and the electroweak $SU(2)\times U(1)$ gauge symmetries with
the one scalar Higgs doublet to break the $SU(2)$ symmetry.
The weak point of the SM is the Higgs mechanism because the resulting quark, lepton and
weak boson masses are derived from free parameters of the Higgs potential and Yukawa couplings.
The masses of the Dirac neutrinos in the SM are the exception because they are well defind and they vanish.
Astonishingly, the only elementary particles that interact only with the massive weak $SU(2)$ bosons
have no masses in the SM.

However, from the measurements of SuperKamiokande in 1998 we know that light neutrinos
have masses and mix their flavours.
This is the first clear signal that the SM has to be substantially redefined.

A more than a decade of measurements at Belle and at LHCb reveal the results not in agreement with
the SM for quantum loop suppressed B meson semileptonic and hadronic decays, as well as for
CP violations with B mesons almost to $5\sigma$ confidence level.

The very precise measurement of the W boson mass by the CDFII attracted a great
attention because their result disagrees with the SM by $7\sigma$ asumming that the mass
of the Higgs boson is $125 GeV$.
The most recent measurement
at the ATLAS with the similar precision agrees well with the SM but disagrees
from the Tevatron CDFII result by more than $5\sigma$. It is important to note that two
measurements are performed at different center of mass energies (2 TeV at CDFII and 13 TeV at ATLAS).

A discovery of the $125 GeV$ resonance in 2012 is later reported as the SM Higgs scalar, though its mass
is too small and the fit of all EW data gives even smaller mass. In addition, there is more
natural interpretation of this resonance.

It is not that only the electroweak part of the SM is faced with problems, but the same is valid
for the standard QCD.

Namely, the huge disagreement between data and theory was observed at Tevatron for charge asymmetry
for t-quarks. The same observable vanishes for the charge symmetric proton-proton LHC collider and
it is necessary to redefine it at the parton level with an order of magnitude smaller SM prediction
very difficult to measure with small experimental errors.

Charged particle multiplicities at the LHC cannot be reproduced by the standard simulation generators
based on the standard QCD - generators underestimate observed multiplicities.

The cosmic rays of the highest energies bring to us the new problem under the name of the muon puzzle.
The center of mass (CM) energies of the observed collisions of cosmic rays with the atmosphere
could be an order of magnitude higher than that of the LHC $13 TeV$. The resulting showers from
hadronic part of interactions produce much more muons than predicted by the simulation generators.

{\bf (B) cosmology - standard LCDM model}

The accepted standard LCDM cosmological model has few distinctive ingredients: 1) it is based
on the general Relativity (GR), 2) inflationary assumption containing a dynamic of some
hypothetical scalar field, 3) dominance of an unknown cold dark matter (CDM) particle and 4) L
stays for the cosmological constant $\Lambda$.

The LCDM assumes the homogeneous and isotropic Friedmann geometry with the initial singularity,
inflaton scalar to fix the total density of the Universe and to define the initial power
spectrum, the existence of CDM particle and the cosmological constant, as well as the
unknown mechanism to break the baryon number.

All the listed cosmological phenomena in LCDM model are supported by the measurements of
the CMB, galaxy correlations, SNae, BBN, weak and strong gravitational lenses, galaxy
rotation curves, etc.

However, few recent years we witness the rise of many disagreements of the LCDM
with the new abundant data from terrestrial and space observatories.

We shall enumerate the following ones: 1) the Hubble tension: more than $5\sigma$ tension
between the measurements of the Hubble constant by the Planck satellite (CMB fluctuations)
vs. local (redshift zero) by Cepheid stars, 2) unconfirmed results of the 21 cm EDGES radio
spectra, 3) $S_{8}$ as a measure of the growth of structures measured by various large scale
surveys does not match the expected LCDM value, 4) surprising results of the high redshift
data from the JWST with overabundant number of mature galaxies that shoul not even exist
at that high redshifts, 5) matter kinematic dipole anomaly - greater than the prediction
of the LCDM.

{\bf (C) gravity and gravity waves}

Very precise measurements and very precise calculations of the binary pulsars have been
considered for a long time as an indirect evidence of the existence of the gravity
waves.

It came out of a no surprise that the enouncement of two LIGO interferometers claiming
the detection of gravity waves attracted enormous attention. I show later that
the lack of any scrutiny and the overall check of the LIGO results, as well as the
ignorance of other analyses, resulted in absurd results of now three interferometer
experiments LIGO, Virgo and KAGRA with impossible black hole masses within the
binary systems.

This issue and the theoretical aspects of the gravity waves I discuss within
the next chapter with the problem resolutions.

\section{New principle and the resolutions of problems in particle physics, cosmology and gravity wave searches}
The attempts to unify forces in physics is based on the assumption that all the forces
(electromagnetic, weak, strong and gravity) are local gauge forces materialized via boson
particles (photon, weak bosons, gluons and graviton). The place of unification is the
Planck scale. Grand Unified Theories, supersymmetric theories in various dimensions or
preon models represent the most frequently studied theories based on the gauge principle.

I assume that the world of elementary particles is able completely to understand with the
noncontractible three dimensional space with a time variable as a necessary parameter to
describe dynamics and kinematics, Minkowski metric and $SU(3)$ conformal family.

The Einstein-Cartan (EC) cosmology based on the Einstein-Cartan nonsingular theory of gravity
is my favourite theory within which I resolve all the main puzzles, tensions and problems
of the LCDM model.

I discuss the issue of the existence of gravity waves within the GR and EC theory.
The interpretation of LIGO, Virgo, KAGRA events as a geophysical
phenomenon is given. The Pulsar Timing Arrays and their data represent an excelent
opportunity to estimate the magnitude of the vorticity of the Universe.

{\bf (A) particle physics}

Since the Higgs mechanism does not solve the problem of the weak bosons and fermion masses,
I introduce the new mechanism - noncontractible space \cite{Palle1}.

Fixing the ultraviolet (UV) cut-off (minimal distance) by the Wick's theorem one can
study now well defined Dyson-Schwinger and Bethe-Salpeter equations, the only essential
equations in the quantum field theory of particle physics. The heaviest particles are
zero node solutions. One accepts as physical solutions only those which develop
mass singularities in timelike domain when interacting with electroweak bosons
\cite{Palle1}. I found zero and one node solutions in the Abelian test model for
an illustration \cite{Palle2}.

The second important ingredient of my theory is the cancellation of the $SU(2)$
global anomalies induced by the massive weak bosons and massive fermions \cite{Palle1}.
The resulting relation between the Weinberg angle and the mixing angles of quarks (leptons)
should be a check of the hidden supersymmetry (symmetry between boson and fermion degrees
of freedom) of my theory. If the inverted mass hierarchy will be experimentally established for
light neutrinos, one has to put the minus sign in front of the corresponding mixing angle
\cite{Palle3}.

The whole algebraic picture of the $SU(3)$ conformal embedding of the SM interactions
\cite{Palle1} is unique and does not allow for new interactions. The fundamental distinction
of my theory from the SM lies in the lepton and scalar sectors: (1) my theory contains
three light and three heavy Majorana neutrinos and (2) instead of the Higgs boson my theory
has unphysical zero norm massless scalar $\zeta$ boson.

One or even two of the heavy Majorana neutrinos are good candidates for cold dark matter
particles \cite{Palle4}. It is not excluded that the HESS source \cite{HESS} in the center of
our galaxy comes from the annihilation of heavy neutrinos.

The mixing matrix of Majorana neutrinos can contain three CP violating phases, however
in my theory the lepton mixing matrix is fixed by Dirac leptons \cite{Palle1}, therefore
only one CP violating phase is present in the final mixing matrix of Majorana neutrinos.
The theory preserves L-B number but the explicit violation of the lepton number by
heavy Majorana neutrinos causes the inevitable violation of the baryon number of the Universe
\cite{Fukugita}.

The lepton sector of my theory is essential to resolve three important problems
of cosmology: (1) light Majorana neutrinos as a hot dark matter particles,
(2) heavy Majorana neutrinos as a cold dark matter particles, (3) broken baryon and lepton
numbers of the Universe.

Let us turn to the electroweak and QCD processes where one can expect considerable
deviation from the SM. The relativity effects are large when velocities are
close to the velocity of light. The quantum effects enter into calculation when the exchanged
energies are small and comparable to the quantum of energy defined by the Planck constant.

We expect that the electroweak quantum loops are measurably modified in my theory when the
heaviest t-quarks are present in the loops. This is precisely what I calculated for
B meson leptonic, semileptonic and hadronic loop suppressed processes
\cite{Palle5,Palle6}. The recent measurements by LHCb and Belle II confirm the deviations.

I should highlight the SM processes where the mass of the exchanged Higgs scalar plays
the essential numerical dependence. My theory contains instead the massless $\zeta$ particle.
The longitudinally polarized same-sign WW scattering is the example of such a process.
The reader can see considerable difference between the differential cross sections of the SM and my theory
\cite{Palle7} with experimental preference to my theory. The three and four t quark productions
with the tree level Higgs exchange or the two Higgs boson productions
could also be a decisive checks of the symmetry breaking
mechanism at the high luminosity LHC.

The QCD with the UV cut-off indicates the enhanced strong coupling at smaller distances
\cite{Palle8} inducing the modifications of the parton distribution functions (PDF).
The t quark charge asymmetry at charge asymmetric Tevatron collider represents the
observable proportional to QCD quantum loop with the observed deviation from the SM but
in accordance with my theory \cite{Palle9}. The cosmic rays with an order of magnitude
larger CM energies striking the atmosphere show much greater production of muons (muon puzzle)
for hadronic part of showers. The larger strong coupling of my theory (note that the deviation from the
standard QCD coupling is logarithmic) can solve this puzzle.

Let us finally consider very interesting measurements of the W boson mass by CDFII (Tavatron)
and ATLAS (LHC).
They are of comparable accuracy but they differ from each other by more than $5\sigma$, but the CDFII result deviates
by $7\sigma$ from the SM value.
I show in my paper \cite{Palle3} that the extrapolation of measured $M_{W}$ to smaller CM energies,
where no disparities between PDFs are expected, my theory perfectly fits to the extrapolated value
of $M_{W}$.

The forthcoming measurements will show whether $125 GeV$ scalar resonance is the Higgs scalar or
the QCD bound state as a mixture of gluonium and toponium \cite{Cea,Palle10}.

{\bf (B) cosmology}

The standard LCDM cosmology based on the GR is a singular theory. Trautman \cite{Trautman} showed
that under standard assumptions on matter, the Einstein-Cartan theory of gravity with
a repulsive force of torsion can avoid the zero distance singularity.

We learned from the particle physics that the solutions in the spacelike domain of Minkowski spacetime
define particle masses in timelike domain. I assume similarly that the mass density of the Universe
can be fixed at infinity \cite{Palle11} using the Penrose conformal technique \cite{Penrose}.
A contribution of torsion to the effective energy-momentum tensor at infinity and its vicinity
(primarily its quadratic term) scales with the cosmic scale factor as $a^{-3}$ \cite{Palle12}, just like
the matter density. The simple algebra gives $\Omega_{m}\simeq 2,\ \Omega_{torsion}\simeq -1,\
\Omega_{\Lambda}\simeq 0$, if vorticity and acceleration parameters are very small
\cite{Palle13,Palle14,Davor1,Davor2}.

Note that the torsion terms give effectively allways a negative contribution since it represents
the rotational degrees of freedom in the linear affine connection of the Riemann-Cartan spacetime.

The primordial density fluctuations can be initiated due to quantum flucutuations of the spin density
of matter \cite{Palle15} at the universal scale given by the weak interactions ($\Lambda_{UV}$).
The heavy Majorana neutrinos as the heaviest particles in my theory (${\cal O}(100 TeV)$), which are
comologically stable $\tau_{N_{i}} > \tau_{Universe}$ \cite{Palle4}, perfectly fit into
the EC cosmology connecting the minimal distance, spin and torsion, number density and the masses
of heavy neutrinos \cite{Palle16}. Therefore we can call these particles the primordial particles
of the Universe.

The ultimate example how local structure of spacetime and its physical content of elementary
particle physics induces the asymmetry within the EC cosmology of Riemann-Cartan spacetime
is hiding in the relation of the weak interaction chirality to the EC cosmic vorticity chirality
\cite{Palle17}. They have allways opposite sign irrelevant of our definition of the coordinate
system.

The recent spectacular observations of the JWST confirm my calculations within the EC cosmology
predicting much larger number of galaxies and more rapid growth of galaxies and black holes at
high redshifts \cite{Palle12,Palle16}.

It is not a surprise that the Hubble constant evaluations from the Planck CMB analyses and from
local Cepheid stars differ considerably. I should like to observe that even the Planck CMB fluctuation
TT spectrum is not consistent with LCDM for all multipoles: reported cosmic observables are inferred from
multipoles $l > 20$. Low mutipoles $l < 20$ are problematic, especially the quadrupole
(low mutipoles describe large angles).
One can observe that the standard LCDM fails at high redshifts. The effective mass density at
lower redshifts for EC cosmology is smaller than $\Omega_{m}\simeq 2$ because of the
repulsive force of torsion (angular momentum) of the nonlinear structures \cite{Palle12,Palle16}.
Since the CMB fluctuations are very small ${\cal O}(10^{-5})-{\cal O}(10^{-3})$, the thorough
analysis within Riemann-Cartan geometry requires the inclusion of vorticity and acceleration
functions. Note that vorticity is induced by spin densities of light neutrinos \cite{Palle17}
and the EC equations of motion. All these phenomena are absent in GR and Friedmann cosmic models.
The future measurements of the redshift drift should clearly reveal what is the real
Hubble function $H(z)$ \cite{Palle16}.

A new and probably the most decisive problem for the LCDM model appears to be a matter
kinematic dipole anomaly. I show that the EC cosmology, unlike LCDM model, reproduces without any problem
much larger kinematic dipole of matter in comparison with the CMB kinematic dipole \cite{Palle18}.
Matter fields (baryons, light and heavy neutrinos) interact with torsion in EC equations
of motion. This is not the case for massless CMB photons.

I should mention interesting estimates of the CMB polarization from the Planck CMB data \cite{Palle14}.
The measured rotation of the polarization vector is a result of the rotating Universe within
the EC cosmology. My relation connects the vorticity of the Universe $\frac{\omega_{0}}{H_{0}}\simeq{\cal O}(10^{-5})$,
angle of rotation of the polarization vector and
the redshift of decoupling \cite{Palle14}.

One can additionally point out that the primordial vorticity inevitably generates
the primordial magnetic fields \cite{Harrison}.

{\bf (C) gravity and gravity waves}

I show \cite{Palle19} that very precise measurements of binary pulsars, considered to be as an
indirect evidence of gravity waves, are completely explainable with the Landau-Lifshitz
exact local conservation equation derived from the Einstein-Hilbert field equations and Bianchi
identity. The loss of the kinetic energy is exactly compensated by the gain of the potential
energy. This is valid as well as in the EC gravity  \cite{Palle19} due to only algebraic
relation between torsion and spin (angular momentum).

Ten years ago LIGO collaboration claimed direct detection of gravity waves.
Only one month later I wrote a paper where the alternative explanation was suggested:
the gravitational action of the ocean tidal bulges on the LIGO interferometers \cite{Palle20}.
Because the mass of the ocean tidal bulges depends on the local depth of the ocean,
the largest gravity action comes from the bulges at the ocean trenches \cite{Palle19}.

The group from Niels Bohr Institute \cite{Bohr} found the following occurrences in the public LIGO data:
roughly one hour before and one hour after the "gravity wave" event they found
backgound events strongly correlated with the "gravity wave" event. Their analysis is in
an ideal conformity with my explanation of the "gravity events" as a geophysical phenomenon.

It is not a wonder that the improved LIGO, Virgo and KAGRA interferometers found
in the past decade few hundreds more "gravity wave" events. Some events refer to the binary
systems with the mass of the black holes in disagreement with the GR. There are no compelling
observed astrophysical processes that should be connected to the detected "gravity wave" events.

I stress in \cite{Palle19} that there is no theoretical proof of the existence of
gravity wave equation, namely the metric satisfies nonlinear integral equation \cite{Sciama}.
Integral equations are not the wave equations.
One can make a linear
approximation only if the existence of the exact gravity wave equation is proved.

{\bf (D) quantum mechanics in noncontractible space}

We know how to treat noncontractible space in the quantum field theory.
I show in \cite{Palle21} that the effect of the noncontractible space in quantum mechanics
can be reduced to the holonomy of the Berry's adiabatic phase resulting in the broadening
of the spectral lines.

\section{Examples of the popular but unphysical concepts}

{\bf Magnetic monopoles}

Dirac proposed the existence of the additional degrees of freedom consisting of the
magnetic poles. I show \cite{Palle1} that the conformal $SU(3)$ unification of
the SM forces $SU(3)_{c}\times SU(2)_{L}\times U(1)$ has no place for the additional
degrees of freedom since $8\times 6 = 8\times 4 + 3\times 4 + 4$.
\newline
\newline
\newline
{\bf Instantons and axions}

Instantons are solutions of the four dimensional Euclidean space. The local physical space
is Minkowski spacetime. A time is a necessary variable to describe kinematics and dynamics
of classical and quantum fields. The time ordered product of fields is the essential concept
in our causal field theory.
Timelike, lightlike and spacelike intervals have important roles in the field theories.
The definite metric of the four dimensional Euclidean space
is therefore responsible for the appearance of the CP violating terms in Minkowski spacetime.
The axion (with mass and couplings to be free parameters) is introduced to save this problem of QCD.

I conclude that instantons (as well as sphalerons) together with the axion are completely
irrelevant and unphysical concepts in our physical Minkowski spacetime.
\newline
\newline
{\bf Hawking radiation of black holes}

Hawking started his analysis with the following "wave equation": \\
$g^{\mu\nu}\nabla_{\mu}\nabla_{\nu}\phi=0$,
where $g_{\mu\nu}=metric,\ \nabla_{\mu}=covariant\ derivative,\ \phi=scalar\ quantum\ field$.
This is ad hoc equation and it is not derived from the Hilbert action. It is a construct in analogy
with the local gauge theories, such as the spinor currents coupled to the external electromagnetic
field.

The scattering matrix of the Hawking "wave equation" is illdefined \cite{Hawking} and it is not
a surprise that on the one side it has the "information loss paradox" and on the other side
the black hole radiates thermally and then vanishes. There is no cure for the Hawking "wave equation".
The only correct way how to relate local physical quantities in GR or in EC gravity to
the spacetime quantities is to mediate effective energy-momentum tensor with tetrads $h^{a}_{\mu}$:
$T^{eff}_{\mu\nu}=h^{a}_{\mu}h^{b}_{\nu}T^{eff}_{a b},\ g_{\mu\nu}=\eta_{a b}h^{a}_{\mu}h^{b}_{\nu},
\ \eta_{a b}=Minkowski\ metric$.

I conclude that the Hawking radiation of black holes is illdefind and unphysical process.

\section{Conclusion}

Mathematical consistency of my considerations in particle physics, cosmology and
gravity and the ability to resolve large list of problems and puzzles in phenomenology of
the local (particle physics) and global (cosmology) structure  of spacetime leads to the
conclusion that my theory of noncontractible space is a law of nature as a third (after quantum
and relativity) missing principle.
\newline
\newline
{\bf Appendix - list of some formulas derived in my papers}
\newline
\begin{eqnarray}
\Lambda_{UV} = \frac{\hbar}{c d}=\frac{\pi}{\sqrt{6}}\frac{2}{g} M_{W},\
 e = g \sin \Theta_{W},\ \cos \Theta_{W} = \frac{M_{W}}{M_{Z}}
\end{eqnarray}
\newline
\begin{eqnarray}
\Theta_{W}=2 (\Theta_{12}^{D}+\Theta_{23}^{D}+\Theta_{31}^{D})
\end{eqnarray}
\begin{eqnarray*}
inverted\ masses: \left( \begin{array}{cc}
\cos \Theta_{ij} & \sin \Theta_{ij} \\
-\sin \Theta_{ij}  & \cos \Theta_{ij} \end{array}
\right)
\left( \begin{array}{c}
u_{i} \\
u_{j} \end{array}
\right)
\Leftrightarrow
\left( \begin{array}{cc}
\cos \Theta_{ij} & -\sin \Theta_{ij} \\
\sin \Theta_{ij}  & \cos \Theta_{ij} \end{array}
\right)
\left( \begin{array}{c}
u_{j} \\
u_{i} \end{array}
\right)  \nonumber \\
\end{eqnarray*}
\begin{eqnarray}
\Rightarrow \Theta^{M}_{ij} =\Theta^{exp}_{ij} = -\Theta^{D}_{ij}
\end{eqnarray}
\newline
\begin{eqnarray}
&&\dot{R}(R_{min})=0\ \Rightarrow \kappa \rho_{\gamma}(R_{min})=\kappa^{2}\frac{\hbar^{2}}{4}n^{2}(R_{min}) \\
&&\Rightarrow n_{0}\simeq \frac{2 R_{min}}{R_{0}\hbar}(\frac{\rho_{\gamma,0}}{\kappa})^{1/2},\
d\simeq 2 R_{min},\ R_{0}\simeq c/H_{0},\ \Omega_{CDM}\simeq 2 \nonumber \\
&&\Rightarrow n_{0}\simeq {\cal O}(10^{-10} cm^{-3})
\Rightarrow m_{CDM}=\rho_{CDM,0}/n_{0}\simeq {\cal O}(100 TeV) \nonumber
\end{eqnarray}
\newline
\begin{eqnarray}
\lim_{R \rightarrow \infty}\frac{\rho_{m}(R)}{\rho_{crit}(R)} = 2
\end{eqnarray}
\newline
\begin{eqnarray}
\chi(weak\ interactions)+\chi(vorticity\ of\ the\ Universe)= -1 +1 = 0
\end{eqnarray}
\newline
\begin{eqnarray}
\beta (rotation\ angle\ of\ CMB\ polarization\ vector)=\frac{\omega_{0}}{H_{0}}z_{dec}
\end{eqnarray}

\end{document}